\documentclass[preprintnumbers, twocolumn,nofootinbib]{revtex4-1}
\usepackage{graphicx}
\usepackage{dcolumn}
\usepackage{bm}
\usepackage{amsmath}
\usepackage{amsfonts} 
\usepackage{latexsym}
\usepackage{bbm}
\usepackage{color}
\newcommand{\non}{\nonumber}

\begin{document}

\preprint{IPMU15-0179}

\title{Toward verification of electroweak baryogenesis by electric dipole moments}

\author{Kaori Fuyuto$^{1}$}%
\email{fuyuto@th.phys.nagoya-u.ac.jp}
\author{Junji Hisano$^{2,1,3}$}%
\email{hisano@eken.phys.nagoya-u.ac.jp}
\author{Eibun Senaha$^{4}$}%
\email{senaha@ncu.edu.tw}
\affiliation{$^1$Department of Physics, Nagoya University, Nagoya 464-8602, Japan}
\affiliation{$^2$Kobayashi-Maskawa Institute for the Origin of Particles and the Universe, Nagoya University, Nagoya 464-8602, Japan}
\affiliation{$^3$Kavli IPMU (WPI), University of Tokyo, Kashiwa, Chiba 277-8584, Japan}
\affiliation{$^4$Department of Physics and Center for Mathematics and Theoretical Physics, National Central University, Taoyuan, 32001, Taiwan}
\bigskip

\date{\today}

\begin{abstract}
We study general aspects of the CP-violating effects on the baryon asymmetry of the Universe (BAU)
and electric dipole moments (EDMs) in models extended 
by an extra Higgs doublet and a singlet, together with electroweak-interacting fermions.
In particular, the emphasis is on the structure of the CP-violating interactions 
and dependences of the BAU and EDMs on masses of the relevant particles.
In a concrete mode, we investigate a relationship between the BAU and the electron EDM
for a typical parameter set. As long as the BAU-related CP violation predominantly 
exists, the electron EDM has a strong power in probing electroweak baryogenesis. 
However, once a BAU-unrelated CP violation comes into play,
the direct correlation between the BAU and electron EDM can be lost.
Even in such a case, we point out that verifiability of the scenario still remains with the help of Higgs physics.
\end{abstract}

\pacs{Valid PACS appear here}

\maketitle

\section{Introduction}
The particle content of the standard model (SM) has been completed
by the discovery of the 125 GeV Higgs boson 
at the Large Hadron Collider (LHC)~\cite{h125_discovery}.
So far, there is no clear signal beyond the SM in laboratory experiments.
Nevertheless, the cosmological problems such as 
the origin of the baryon asymmetry of the Universe (BAU) and 
identification of the cold dark matter still remain unsolved within the SM.

One of the mechanisms for generating the BAU is electroweak baryogenesis (EWBG)~\cite{ewbg}.
In this scenario, the BAU arises during the electroweak phase transition (EWPT),
and its feasibility depends on properties of models at the GeV/TeV scales.
From the viewpoint of the testability, 
EWBG is the first scenario that is verified or falsified by the ongoing and upcoming
experiments, among others.
As is well known, the SM has the two drawbacks that prevent it from generating the BAU:
absence of both a strong first-order EWPT~\cite{lattice} 
and a sufficient amount of CP violation~\cite{KMphase}.
Supersymmetric (SUSY) models may naturally solve those issues simultaneously.
For example, in the the minimal SUSY SM model (MSSM), a light scalar top (stop) could induce 
the strong first-order EWPT, and the fermionic superpartners 
provide the substantial amount of CP violation.
However, it turns out that the light stop scenario in the MSSM 
is not consistent with the LHC Run 1 data 
such as the Higgs signal strengths and the direct stop searches~\cite{MSSM-EWBG_LHCtension}. 
Given this fact, the colored particles may no longer the candidates for archiving 
the strong first-order EWPT.
Therefore, whatever a UV theory might be, the possibility of EWBG can be investigated
in the framework of an effective field theory of non-colored particles after integrating out 
irrelevant heavy degrees of freedom, {\it i.e.},
\begin{align}
\mbox{UV theories} \supset\mbox{multi-Higgs}+\mbox{EW-interacting fermions}.
\end{align}

Experiments that are most sensitive to the CP violation are measurements 
of the electric dipole moments (EDMs) of electron, neutron and atoms etc. 
Clarifying relationships between the BAU-related CP violations and the EDMs
are indispensable for the test of the EWBG scenario.
In some analyses in the literature, the CP-violating effect is incorporated 
by higher dimensional operators assuming only one Higgs doublet
and by which the BAU is evaluated. 
In such a case, the CP-violating effects peculiar to the finite temperature,
such as a resonant enhancement pointed out in Ref.~\cite{CTP}, are missing,
which drastically changes the correlation between the BAU and EDM.

In this Letter, we clarify similarities and differences between the BAU-related CP violation
and the EDM-related one with particular emphasis on 
the structure of the interactions and the mass dependences of the relevant particles. 
As an illustration, we consider a framework in which the Higgs sector is augmented 
by an additional Higgs doublet and a singlet, and in addition, SU$(2)_L$ doublet fermions 
and singlet fermion are introduced to accommodate CP violation for baryogenesis.
In our setup, the structure of the CP-violating interactions are more generic than those
in SUSY models.
We evaluate the CP-violating source term for the BAU 
in the closed-time-path formalism and relate it with the electron EDM.
The correlation between the two CP-violating quantities is elucidated
as functions of the EW-interacting fermion masses.

As a specific example, we consider a next-to-MSSM-like model and 
work out the relationship between the BAU and electron EDM.  
It is found that the electron EDM is the useful probe of the baryogenesis favored region 
as long as the BAU-related CP violation predominately exists in the model. 
However, there is a case in which a BAU-unrelated CP violation, if it exists, alters the intimate connection
between the BAU and EDM, which makes it difficult to test EWBG 
via the electron EDM experiment only.
Nevertheless, such a specific case is possible only in the case that the doublet-singlet Higgs boson
mixing exists, which is needed for a tree-potential-driven strong first-order EWPT, 
and thus still testable in combination with Higgs physics.

%
\section{General aspects of CP-violating effects on the BAU and EDMs}\label{sec:BAUCPV}
%
Before going to present our model, 
we here give a simple but rather generic argument about the relationship between
the BAU-related CP violation and EDM.
For illustrative purposes, we consider the framework in which two Higgs doublets 
and two species of EW-interacting fermions (denoted as $\psi_{i,j}$) are present. 
For definite, $\psi_i$ is assumed to be Dirac fermion and $\psi_j$ Majorana fermion.
This setup applies to the bino-driven EWBG in the MSSM~\cite{Li:2008ez}, 
the singlino-driven EWBG in the next-to-MSSM~\cite{Cheung:2012pg} and the $Z'$ino-driven EWBG 
in the U$(1)'$-MSSM~\cite{Senaha:2013kda} in proper limits.
We expect that the following discussion would hold in other cases by making
an appropriate translation.

Let us parameterize the relevant interactions as
\begin{align}
\mathcal{L}= \frac{1}{\sqrt{2}}\bar{\psi}_i\big(c_Lv_a P_L+c_Rv_b P_R\big)\psi_j+{\rm h.c.},
\label{inos_int}
\end{align}
where $v_{a,b}\;(a, b=1,2)$ 
denote the Higgs vacuum expectation values (VEVs), and $c_{L,R}$ are the complex parameters.
With this Lagrangian, we evaluate the source terms in the diffusion equation of $\psi_i$
in the closed-time-path formalism~\cite{CTP}. The vector current of $\psi_i$ has the form
\begin{align}
\partial_\mu j^\mu_{\psi_i}
= S_{\psi_i},
\end{align}
where only the CP-violating source term is shown on the right-hand side.
\begin{figure}[t]
\begin{center}
\includegraphics[width=5.5cm]{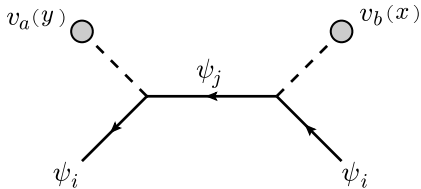} 
\end{center}
\caption{A representative scattering process of $\psi_i$ with the Higgs bubble walls,
which leads to a dominant CP-violating source term for the BAU.}
\label{fig:self_fer}
\end{figure}
In a VEV insertion approximation~\cite{CTP}, $S_{\psi_i}$ to leading order is induced 
by the process shown in Fig.~\ref{fig:self_fer}, which is cast into the form
\begin{align}
S_{\psi_i}(X) 
&= \kappa_S\cdot 2m_{i}m_{j}{\rm Im}(c_Lc_R^*)v^2(X)\dot{\beta}(X)\mathcal{I}_{ji}^f \nonumber\\
&\equiv C_{\rm BAU}{\rm Im}(c_Lc_R^*),
\label{Scpv}
\end{align}
where $\kappa_S=+1$ for $(a, b)=(2,1)$, 
$\kappa_S=-1$ for $(a, b)=(1,2)$
and $\kappa_S=0$ for $(a, b)=(1,1),\;(2,2)$.
$m_{i,j}$ are the masses of $\psi_{i,j}$, $\dot{\beta}(X)$ is the time derivative of 
$\beta(X)=\tan^{-1}(v_2(X)/v_1(X))$, and $\mathcal{I}_{ji}^f$ denotes a thermal function
as will be given below.
One can see that $S_{\psi_i}(X)$ would vanish not only for ${\rm Im}(c_Lc_R^*)=0$ but also
the cases in which one of the following condition is fulfilled:
$ v(X) = 0$, $\dot{\beta}(X) = 0$ and $\mathcal{I}_{ji}^f = 0$.
Since the EWPT is of first order, the Higgs VEVs depend on a spacetime variable $X$,
and the profiles of which can be determined by static bubble configurations at a
nucleation temperature. 
In most cases, the shapes of $v(X)$ and $\beta(X)$ 
would be approximated by kink-type configurations,
so the $\dot{\beta}(X)$ is proportional to a variation of $\beta(X)$ along the line
connecting broken and symmetric phases.
In the MSSM, $\dot{\beta}(X)$ roughly scales as $1/m_A^2$~\cite{Moreno:1998bq}, 
where $m_A$ is the CP-odd Higgs boson mass,
which implies that $S_{\psi_i}(X)$ in Eq.~(\ref{Scpv}) would completely disappear
if the Higgs sector is composed of only one Higgs doublet, as already indicated 
in the case of $\kappa_S=0$.
From this argument, it is expected that the presence of the extra Higgs boson with a nonzero VEV
may be essential for successful EWBG, 
regardless of the strong first-order EWPT realization.
Here, it should be reminded that there is another type of the source term that 
is not suppressed in the large $m_A$ limit, 
which may appear as a higher order correction 
to the approximation we have made here (see, {\it e.g.}, Refs.~\cite{Carena:2000id,Carena:2002ss}).
As long as the BAU is explained by a resonant enhancement, which is indeed the case
in our analysis, such a source term would not play a central role.

The behavior of the thermal function $\mathcal{I}_{ji}^f$ is somewhat complicated, 
and in some specific region it is strongly governed by the finite temperature physics. 
The explicit form of $\mathcal{I}_{ji}^f$ is~\cite{CTP}
\begin{align}
\mathcal{I}_{ji}^f&= \int_k\; \frac{k^2}
	{\omega_j\omega_i} 
	\Big[
	\Big\{\big(1-2{\rm Re}(n_i)\big)I_{ji}+(i\leftrightarrow j)\Big\}  \nonumber\\
&\hspace{2.5cm}
	-2\big({\rm Im}(n_j)+{\rm Im}(n_i)\big)G_{ji}
\Big],
\end{align}
where $\int_k=\int_0^\infty dk/(4\pi^2)$, 
$n_{i}=1/(e^{(\omega_{i}-i\Gamma_{i})/T}+1)$, $\omega_i=\sqrt{k^2+m_i^2}$, 
with $\Gamma_i$ being the thermal widths of $\psi_{i}$. 
Here, $I_{ij}$ and $G_{ij}$ are respectively expressed by
\begin{align}
I_{ij}
&= \Gamma_+
\left[
	\frac{\omega_+}{(\omega_+^2+\Gamma_+^2)^2}
	+\frac{\omega_-}{(\omega_-^2+\Gamma_+^2)^2}
\right], \\
G_{ij}&= \frac{1}{2}
\left[
	\frac{\omega_+^2-\Gamma_+^2}{(\omega_+^2+\Gamma_+^2)^2}
	-\frac{\omega_-^2-\Gamma_+^2}{(\omega_-^2+\Gamma_+^2)^2}
\right],
\end{align}
where $\omega_\pm = \omega_i\pm \omega_j$ and $\Gamma_+=\Gamma_i+\Gamma_j$.
One can see that $\mathcal{I}_{ji}^f$ vanishes if $\Gamma_i=\Gamma_j=0$.
Since $\Gamma_{i,j}\simeq gT$, where $g$ represents a typical coupling in a model
and $T$ a temperature, $S_{\psi_i}(X)$ first emerges to order of $\mathcal{O}(g^4)$
assuming $|c_L|=|c_R|\simeq g$.

As is well known, $S_{\psi_i}$ has a resonant enhancement at $m_i=m_j$,
the behavior of which comes from $G_{ij}$.
Since $\omega_{i,j}\gg \Gamma_{i,j}$, one may approximate $G_{ij}$ as
\begin{align}
G_{ij} \simeq  -\frac{1}{2}\frac{\omega_-^2-\Gamma_+^2}{(\omega_-^2+\Gamma_+^2)^2}
	+\mathcal{O}\left(\frac{1}{\omega_+^2}\right).
\end{align}
One can see that $G_{ij}$ has a peak at $\omega_- =0$,
which can yield the dominant source for the BAU.

\begin{figure}[t]
\begin{center}
\begin{tabular}{cc}
\includegraphics[width=3.3cm]{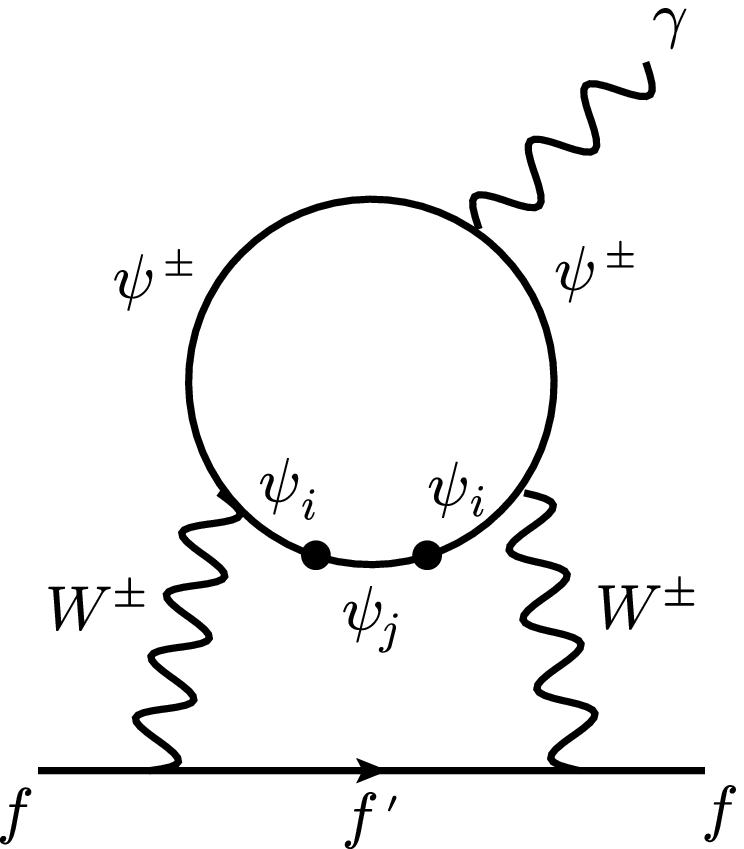} 
\includegraphics[width=3.3cm]{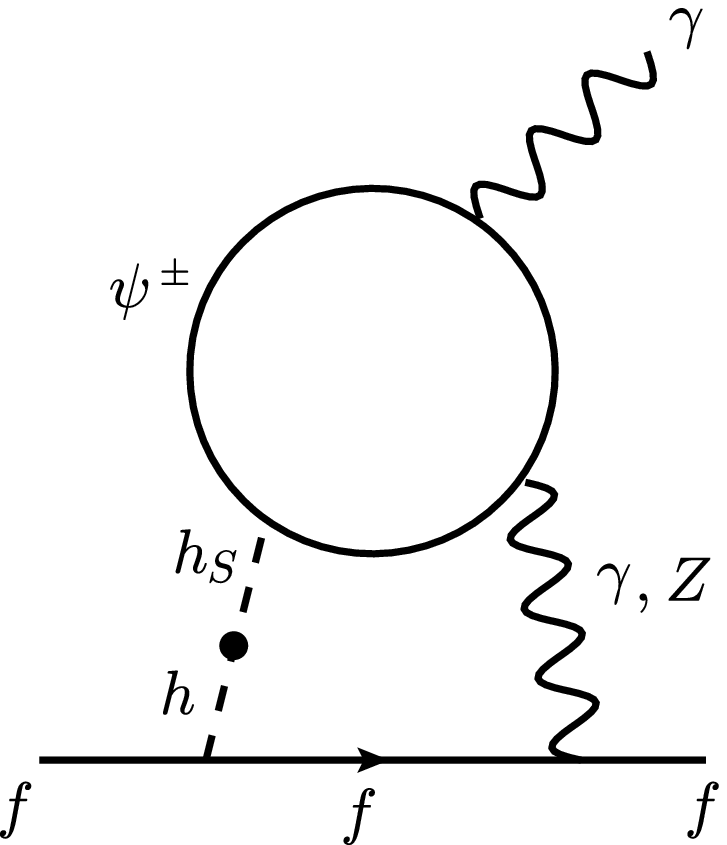} 
\end{tabular}
\end{center}
\caption{Two-loop Barr-Zee diagrams induced by the BAU-related CP violation (Left) 
and the BAU-unrelated one (Right). 
The blobs indicate the mass insertions. Here, $h$ and $h_S$ are the Higgs bosons coming from the doublet and singlet, respectively. 
The size of the $h$-$h_S$ mixing is intimately related to strength of the strong first-order EWPT
in the tree potential driven scenario.}
\label{fig:bar_zee}
\end{figure}

We now study the impact of ${\rm Im}(c_Lc_R^*)$ on the EDM.
Since the new fermions have the EW charges, the following interactions exist.
\begin{align}
\mathcal{L}
& =
\frac{g_2}{\sqrt{2}}
\left(
	\overline{\psi^+}\gamma^\mu\psi_iW_\mu^+ 
	+\overline{\psi}_i\gamma^\mu\psi^+W_\mu^-
\right)
	-e\overline{\psi^+}\gamma^\mu\psi^+A_\mu,
\end{align}
where $\psi^\pm$ denote electrically charged members in the SU$(2)_L$ multiplet fermion.
We assume that $\psi_i$ is the neutral member of the same multiplet.
In this case, the $WW$-mediated Barr-Zee diagram is induced, as shown in Fig.~\ref{fig:bar_zee}.
\footnote{Barr-Zee diagrams involving the heavy Higgs bosons are also generated.
Here, we assume that those Higgs bosons are heavy enough not to 
alter the following discussion drastically. The case without this assumption will be given in~\cite{FHS}.}
The EDM of a fermion $f$ using the mass insertion method is given by
\begin{align}
\frac{d_f^{WW}}{e} 
&= \mp\frac{\alpha_{\rm em}^2}{64\pi^2s_W^4}
\frac{m_f m_{\psi^\pm}m_jv_av_b}{m_W^4}{\rm Im}(c_Lc_{R}^*)F_{WW} \nonumber\\
&\equiv C_{\rm EDM}^{WW}{\rm Im}(c_Lc_{R}^*).
\label{dfWW}
\end{align}
where the negative (positive) sign is the case that $f$ is up-type (down-type) fermion,
$F_{WW}=(f_{WW}(r_i,r_+)-f_{WW}(r_j,r_+))/(m_i^2-m_j^2)$ with $r_i=m_i^2/m_W^2$,
$r_j=m_j^2/m_W^2$ and $r_\pm=m_{\psi^\pm}^2/m_W^2$. 
The explicit form of $f_{WW}$ is given in Ref.~\cite{Ellis:2010xm}.
We emphasize that unlike $S_{\psi_i}(X) $ in Eq.~(\ref{Scpv}), 
Eq.~(\ref{dfWW}) does not vanish for $(a,b)=(1,1)$ or $(2,2)$,
in addition, $d_f^{WW}/e$ is not enhanced at $m_i=m_j$,
which are the prominent differences between the two CP-violating quantities.
One may find that $d_f^{WW}/e \propto m_f m_j/m_i^3$ for $m_i\gg m_j$ and 
$d_f^{WW}/e \propto m_f/(m_im_j)$ for $m_j\gg m_i$,
which signifies another distinct feature of the EDM as discussed below.
In what follows, we confine ourself to the cases of $(a, b)=(2,1)$ and $(1,2)$.

It is worth making a comment on that the mass insertion method used in Eq.~(\ref{dfWW}) 
not only makes it easy to see the relationship 
between the CP-violating source term and the EDM 
but also gives the numerically good approximation.

Eliminating ${\rm Im}(c_Lc_R^*)$ in Eq.~(\ref{Scpv}) using Eq.~(\ref{dfWW}), one finds
\begin{align}
S_{\psi_i} = \frac{C_{\rm BAU}}{C_{\rm EDM}^{WW}}\left(\frac{d_f^{WW}}{e}\right).
\label{Scpv-dfWW}
\end{align}
In order to see the correlation between $S_{\psi_i}$ and $d_f^{WW}/e$ in more detail, we define
\begin{align}
\bar{S}_{\psi_i}=\frac{C_{\rm BAU}}{v^2(X)\dot{\beta}(X)C_{\rm EDM}^{WW}}
\cdot \left(\frac{d_f^{WW}}{e}\right)_{\rm EXP}.\label{Sbar}
\end{align}
In what follows, we consider the electron EDM as 
the experimental constraint, {\it i.e.,} 
$|d_e^{\rm exp}| = 8.7\times 10^{-29}~{e\cdot \rm cm}$~\cite{Baron:2013eja}.
Here, we get rid of $v^2(X)\dot{\beta}(x)$ in $C_{\rm BAU}$ since it is rather model dependent.

\begin{figure}[t]
\center
\includegraphics[width=5cm]{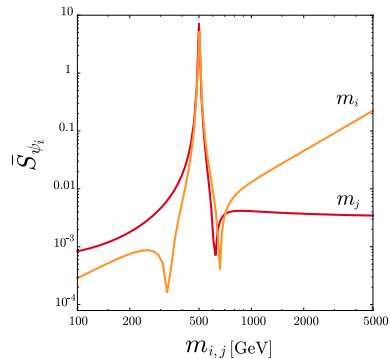} 
\caption{$\bar{S}_{\psi_i}$ as a function of $m_i$ with a fixed $m_j$ and the other away around. We set $\tan\beta=1$ and the fixed mass is $500$ GeV.}
\label{fig:Sbar}
\end{figure}

In Fig.~\ref{fig:Sbar}, 
$\bar{S}_{\psi_i}$ is plotted as a function of $m_i$ with a fixed $m_j$ or the other away around.
As an example, we take $\tan\beta=1$, and the fixed mass is set to $500$ GeV.
As explained above $C_{\rm BAU}$ has a peak at $m_i=m_j$.
However, the decoupling behaviors in the large mass limits are substantially different from each other.
For the varying $m_j$ case, $\bar{S}_{\psi_i}$ becomes more or less flat in the large mass region
while it grows for the varying $m_i$ case.
The latter is due to the rapid suppression of $C_{\rm EDM}^{WW}$ that scales as $m_j/m_i^3$
as mentioned above.
Note that ${\rm Im}(c_Lc_R^*)\gtrsim 1$ for $m_i\gtrsim 1$ TeV since $d_f^{WW}/e$ is fixed. 

Now we move on to discuss a possibility that the aforementioned correlation between 
the CP-violating source term and the EDM is spoiled 
by contamination of BAU-unrelated CP violation.
As delineated below, such a situation can arise when we address the issue 
of the strong first-order EWPT.

The SM Higgs sector has to be extended in such a way that the EWPT is of first order.
There are two representative cases for achieving this:
\begin{itemize}
\item Thermal loop driven case
\item Tree potential driven case
\end{itemize}
For example, the former corresponds to the SM, MSSM 
and a two Higgs doublet model (2HDM) and so on.
In such cases, the cubic-like terms arising from the bosonic thermal loops
play an essential role in inducing the first-order EWPT.
In the latter case, on the other hand, a specific structure of a tree-level Higgs potential
is the dominant source for generating a barrier separating the two degenerate minima 
at a critical temperature. 
One of such an example is the EWPT in the SM 
with a real singlet Higgs boson (rSM)~\cite{rSM,Fuyuto:2014yia}.
In this case, nonzero doublet-singlet Higgs mixing terms are responsible for
the strong first-order EWPT.
Once the singlet Higgs field ($S$) exists, it is conceivable that the following interactions 
may give rise to an extra source for CP violation.
\begin{align}
\mathcal{L}= \overline{\psi^+}(g^S+i\gamma_5g^P)\psi^+S.\label{psi-psi-S}
\end{align}
If the doublet-singlet Higgs mixing is present, 
the Higgs-photon($Z$)-mediated Barr-Zee diagrams 
could be generated, as depicted in Fig.~\ref{fig:bar_zee}. 
In this case, the EDM is the sum of those diagrams, in addition to $d_f^{WW}$.
\begin{align}
\frac{d_f}{e} = \frac{d_f^{WW}}{e}+\frac{d_f^{H\gamma}}{e}+\frac{d_f^{HZ}}{e}.
\end{align}
As far as EWBG is concerned, the new CP-violating phase appearing in Eq.~(\ref{psi-psi-S})
is not directly related to baryogenesis.
Therefore, the linear correlation between the CP-violating source term and the EDM 
in Eq.~(\ref{Scpv-dfWW}) no longer hold.
One of the interesting possibilities is that
if a cancellation among those contributions becomes effective, it is possible for $d_f$ 
to be made highly suppressed but with the nonzero $d_f^{WW}$, 
so the BAU-related CP violation is not constrained by a single 
EDM experiment in this case. 

Nevertheless, one may probe such a parameter space with Higgs physics since
the nonzero doublet-singlet Higgs mixing parameter and $g^{S,P}$ would lead to some deviations 
in the Higgs signal strengths. 
We will explicitly demonstrate this possibility in the next section.

So far, we have exclusively focused on the relationship 
between the CP-violating source term and the EDM.
Here, we comment on the dependence of ${\rm Im}(c_Lc_R^*)$ 
on the baryon number density ($n_B$) briefly.
Under some mild assumptions, one may have
\begin{align}
n_B = \kappa_B\frac{S_{\rm CPV}}{\sqrt{\Gamma_{\rm CPC}}},
\end{align}
where $\kappa_B$ is a coefficient.
$S_{\rm CPV}$ is a CP-violating term arising from
$S_{\psi_i}$ discussed above and
$\Gamma_{\rm CPC}$ a CP-conserving particle changing rate. 
For the latter, for example, the interactions in Eq.~(\ref{inos_int}) induce
\begin{align}
\Gamma_{\psi_i}(X)
&= \frac{1}{T}
\Big[
\big(|c_L|^2v_a^2(X)+|c_R|^2v_b^2(X)\big)\mathcal{F}_{ji} \nonumber\\
&\hspace{1cm}
	+2{\rm Re}\big(c_Lc_R^*\big)v_1(X)v_2(X) m_im_j\mathcal{R}_{ji}\Big],
\end{align}
where $\mathcal{F}_{ji}$ and $\mathcal{R}_{ji}$
are the thermal functions presented in Ref.~\cite{Senaha:2013kda}.
As studied in Ref.~\cite{Lee:2004we}, 
$\Gamma_{\psi_i}$ also has the resonant behavior at $m_i=m_j$,
rendering $n_B$ smaller.
It should be emphasized that a cancellation between the first and second terms 
in $\Gamma_{\psi_i}$ can happen depending on the choice of ${\rm Arg}(c_Lc_R^*)$
and $m_{i,j}$.
Therefore, $n_B$ does not necessarily take its maximal value at ${\rm Arg}(c_Lc_R^*)=\pi/2$ or $-\pi/2$, which may relax the EDM constraint to some extent.

%
%
\section{A model}\label{sec:Model}
\begin{table}[t]
\begin{center}
\begin{tabular}{|c|c|c|}
\hline
particles & SU(3)$_C$ $\times$ SU(2)$_L$ $\times$  U(1)$_Y$ &$Z_2$ \\
\hline\hline
$\Phi_1$ & $({\bf1,~2,~}1/2)$ & $-$\\
$\Phi_2$ & $({\bf1,~2,~}1/2)$ & $+$\\
$S$ & $({\bf1,~1,~}0)$ & $-$\\
\hline
$\tilde{\Phi}_1$ & $({\bf1,~2,~}-1/2)$ & $-$\\
$\tilde{\Phi}_2$ & $({\bf1,~2,~}1/2)$ & $+$\\
$\tilde{S}^0$ & $({\bf1,~1,~}0)$ & $-$\\
\hline
\end{tabular}
\end{center}
\caption{Particle content of the Higgs and the new EW-interacting fermion sectors.}
\label{Charge_assign}
\end{table}
Now, we define our model and give basic ingredients for calculating the BAU and the electron EDM.
The particle content of the Higgs and the new EW-interacting fermion sectors in the model
is shown in Table~\ref{Charge_assign}.
The total Lagrangian is given by
\begin{align}
\mathcal{L}  
&= \mathcal{L}_{\rm 2HDM}+\frac{1}{2}\partial_\mu S\partial^\mu S-V_S-V_{\Phi S}
+\mathcal{L}_{\tilde{\Phi}\tilde{S}}, \nonumber\\
\mathcal{L}_{\tilde{\Phi}\tilde{S}}
&=\sum_{i=1,2}\overline{\tilde{\Phi}}_ii\bar{\sigma}^\mu D_\mu \tilde{\Phi}_i
+\overline{\tilde{S}^0}i\bar{\sigma}^\mu \partial_\mu\tilde{S}^0 \non\\
&\quad -\epsilon_{ab}
\bigg[\sum_{j=1,2}
\left(\tilde{\Phi}_1^ac_{1j}\Phi_j^b+\tilde{\Phi}_2^ac_{2j}(i\tau^2\Phi_j^{b*})\right)\tilde{S}^0 
\nonumber \\
&\hspace{3.5cm} +(\mu+\lambda S)\tilde{\Phi}_1^a\tilde{\Phi}_2^b+{\rm h.c.}
\bigg] \nonumber\\
&\hspace{1.5cm}
+\frac{1}{2}(\mu_{\tilde{S}}+\kappa S)\tilde{S}^0\tilde{S}^0+{\rm h.c.},
\end{align}
where $\tilde{\Phi}_{1,2}$ and $\tilde{S}^0$ are the two-component spinors,
and $\epsilon_{12}=-\epsilon_{21}=+1$. 
As is the case in the MSSM, 
to avoid a lepton flavor violation, we impose a matter parity under which
new EW-interacting fermions are odd and the SM fermions are even.
Furthermore, as in the ordinary 2HDM, 
another $Z_2$ symmetry ($\Phi_1\to -\Phi_1$ and $\Phi_2\to \Phi_2$) 
is enforced to evade tree-level Higgs-mediated flavor-changing-neutral current processes.
Depending on $Z_2$ charge assignments for the fermions, four types of the Yukawa
interactions are possible. However, the following analysis does not depend on those types
since the top Yukawa coupling is the only relevant that is common to all the types.

The Higgs fields are parametrized as
\begin{align}
\Phi_{i=1,2}(x)&=
\begin{pmatrix}
\phi^+_i\\
\phi^0_i
\end{pmatrix}=
\begin{pmatrix}
\phi^+_i\\
\frac{1}{\sqrt{2}}(v_i+h_i(x)+ia_i(x))
\end{pmatrix},\\
S(x)&=v_S+h_S(x),
\end{align} 
where $v_1 = v\cos\beta$, $v_2 = v\sin\beta$ with $v=246$ GeV.


In the following, we consider a rSM-like limit in which $\sin(\beta-\alpha)=1$, 
where $\alpha$ denotes a mixing angle between two CP-even 
Higgs bosons ($h_{1,2}$). 
In this case, only one state (defined as $h$) has the VEV and gives the masses 
of the gauge bosons and fermions.
Since the strong first-order EWPT is assumed to be driven by the tree-Higgs potential, 
the heavy Higgs bosons do not necessarily have the so-called
nondecoupling effect which is needed 
in the thermal loop driven strong first-order EWPT case~\cite{2HDM} .
The detailed comparison between the two cases will be given elsewhere~\cite{FHS}.

Since we have the singlet Higgs boson in this model, $h$ mixes with $h_S$ through
a mixing $\gamma$ as
\begin{align}
\begin{pmatrix}
h \\
h_S
\end{pmatrix}
=
\begin{pmatrix}
c_\gamma & -s_\gamma \\
s_\gamma & c_\gamma
\end{pmatrix}
\begin{pmatrix}
H_1 \\
H_2
\end{pmatrix}.
\end{align}
In our scenario, $H_1$ is the SM-like Higgs boson whose mass is $125$ GeV,
and $H_2$ is the singlet-like Higgs boson which is assumed to be heavier than $H_1$.
Another CP-even Higgs boson originated from the Higgs doublet 
is denoted as $H_3$ which is heavier than $H_2$.

In response to $Z_2$ charges assignments of $\tilde{\Phi}_{1,2}$ and $\tilde{S}$, 
there are several types of the interactions among the new EW-interacting fermions 
and Higgs bosons~\cite{FHS}. Here, we focus on one of them as an example. 
The $Z_2$ charge assignment is listed in Table.~\ref{Charge_assign}.

The relevant interactions among the EW-interacting fermions and Higgs bosons are
\begin{align}
\mathcal{L}^{\rm int}_{\tilde{\Phi}\tilde{S}}
\ni&-\sum_{i=1,2}H_i\overline{\tilde H^+}\left(g^S_{H_i\bar{\tilde H}\tilde{H}}
	+i\gamma_5g^P_{H_i\bar{\tilde H}\tilde{H}}\right)\tilde{H}^+\nonumber\\
%
&+\Big[
\overline{\tilde H^0}\left(c^{\tilde{H}^{0}\tilde{S}}_L\phi^0_2P_L+c^{\tilde{H}^{0}\tilde{S}}_R\phi^0_1P_R\right){\tilde S}+{\rm h.c.}\Big], 
\label{fermion_pote}
\end{align}
where the fermions are expressed in terms of the four-component spinors.
Each coupling is respectively given by
\begin{align}
g^S_{H_{1}\bar{\tilde H}\tilde{H}}&=|\lambda|\cos\phi_{\lambda\tilde{H}}s_\gamma,\quad
g^P_{H_1\bar{\tilde H}\tilde{H}}=-|\lambda|\sin\phi_{\lambda\tilde{H}}s_\gamma,\\
g^S_{H_2\bar{\tilde H}\tilde{H}}&=|\lambda|\cos\phi_{\lambda\tilde{H}}c_\gamma,\quad
g^P_{H_2\bar{\tilde H}\tilde{H}}=-|\lambda|\sin\phi_{\lambda\tilde{H}}c_\gamma,\\
c^{\tilde{H}^{0}\tilde{S}}_L&=-c_{12}e^{-i\phi_{\tilde S}/2},\quad
c^{\tilde{H}^{0}\tilde{S}}_R=c^*_{21}e^{i(\phi_{\tilde H}+\phi_{\tilde S}/2)},
\end{align}
where we have defined $\lambda=|\lambda|e^{i\phi_{\lambda}}$, $\mu+\lambda v_S=|\mu+\lambda v_S|e^{i\phi_{\tilde H}}$, $\mu_{\tilde S}=|\mu_{\tilde S}|e^{i\phi_{\tilde S}}$
and $\phi_{\lambda\tilde{H}} = \phi_{\lambda}-\phi_{\tilde{H}}$.
As discussed in the previous section, the interactions in the second line of Eq.~(\ref{fermion_pote}) 
plays an essential role in generating the CP-violating term that fuels the BAU.
For notational simplicity, we define $\phi=-(\phi_{\tilde H}+\phi_{\tilde S})$ hereafter.

\section{Numerical analysis}\label{sec:Num}
Following a calculation method formulated and developed 
in Refs.~\cite{CTP,Lee:2004we,Huet:1995sh},
we estimate $n_B$ by 
\begin{align}
n_B=\frac{-3\Gamma^{(s)}_B}{2\sqrt{v^2_w+4{\cal R}D_q}}\int^{0}_{-\infty}dz^{\prime}n_L(z^{\prime})e^{-\lambda_-z^{\prime}},
\end{align}
where $\lambda_-=\left[v_w-\sqrt{v^2_w+4{\cal R}D_q}\right]/(2D_q)$,
$\Gamma^{(s)}_B$ is a baryon number changing rate in the symmetric phase,
$v_w$ is a velocity of the bubble wall, $D_q$ is a diffusion constant of the quarks,
and ${\cal R}$ is a relaxation term, which is $(15/4)\Gamma^{(s)}_B$ in our model.
$n_L$ is the total number density of 
all the left-handed quarks and leptons~\cite{Huet:1995sh,Cohen:1994ss}.

Since the EWPT is reduced to that in the rSM, 
we adopt S2 scenario investigated in Ref.~\cite{Fuyuto:2014yia} as a benchmark
in which $m_{H_2}=170$ GeV, $\cos\gamma\simeq0.94$ and $v_C/T_C=206.75~{\rm GeV}/111.76~{\rm GeV}$.
In addition, we take $\tan\beta=1$, $v_w=0.4$, $\Gamma_{\tilde{H}} = 0.025T$, 
$\Gamma_{\tilde{S}} = 0.003T$,
and use an approximation, $\dot{\beta}=v_w\Delta\beta/L_w$ taking $\Delta\beta=0.015$.
Under this assumption, $n_B$ does not depend on $L_w$. 
Moreover, the constant VEV but $v_C/2$ is used in calculating $n_B$,
which may give a simple approximation of kink-type VEV~\cite{FHS}.
For the heavy Higgs boson masses, we set 400 GeV, and for a softly $Z_2$ broken 
mass, which is a mixing mass between $\Phi_1$ and $\Phi_2$, 250 GeV is taken. 
For the other parameters, we refer to the values adopted in Ref.~\cite{Senaha:2013kda}.
In the following, the electron EDM is calculated in the mass eigenbasis of the neutral fermions rather than the mass insertion method, 
although the both are not much numerically different.

%

\begin{figure}[t]
\center
\includegraphics[width=4.8cm]{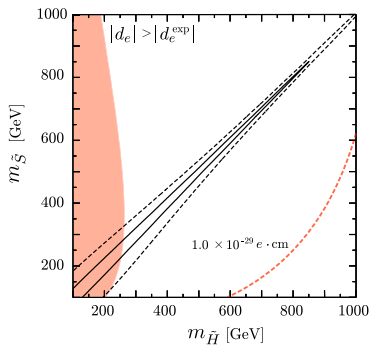} 
\caption{The contours of $Y_B/Y_B^{\rm obs}$ and $|d_e|$ in the $(m_{\tilde{H}}, m_{\tilde{S}})$ plane. The region colored in orange is excluded by the current experimental limit of the electron EDM, and the orange dashed line corresponds to $|d_e|=1.0\times 10^{-29}~{e\cdot \rm cm}$. The black solid and dashed lines represent $Y_B/Y_B^{\rm obs}=1$ and $0.1$, respectively. We set $|c^{\tilde{H}^{0}\tilde{S}}_L|=|c^{\tilde{H}^{0}\tilde{S}}_R|=0.42$ and $\phi=225^\circ$.}
\label{fig:bau03_225}
\end{figure}
We first present the case where the electron EDM is induced by only the $WW$-mediated Barr-Zee diagram. In Fig.~\ref{fig:bau03_225}, contours of $Y_B/Y_B^{\rm obs}$ and $|d_e|$ are shown in the $(m_{\tilde{H}}, m_{\tilde{S}})$ plane. 
We take $|c^{\tilde{H}^{0}\tilde{S}}_L|=|c^{\tilde{H}^{0}\tilde{S}}_R|=0.42,~\phi=225^\circ$ and $|\lambda|=0$. 
Here, $\phi$ is chosen in such a way that the cancellation
in $\Gamma_{\rm CPC}$
is effective.
In this figure, the orange region is excluded by the current experimental limit of the electron EDM, 
$|d_e^{\rm exp}|<8.7\times 10^{-29}~{e\cdot \rm cm}$, and the dashed line corresponds to 
$|d_e|=1.0\times10^{-29}~{e\cdot \rm cm}$ which is reachable by the future experiments~\cite{future_eEDM}. 
The black solid (dashed) line indicates $Y_B/Y_B^{\rm obs}=1~(0.1)$. 
One can see that $|d_e|$ gets rapidly suppressed as $m_{\tilde{H}}$ 
increases but does not in the large $m_{\tilde{S}}$ case, as discussed in Sec.~\ref{sec:BAUCPV}.
Furthermore, the BAU is sufficiently generated if $m_{\tilde{H}}\simeq m_{\tilde{S}}$ due
to the resonant effect.
Our result shows that the successful EWBG region would be entirely verified 
by the future experiments of the electron EDM 
even if the BAU calculated here is underestimated by a factor of 10 or even more
due to lack of precise knowledge of the bubble profiles etc.

\begin{figure}[t]
\center
\includegraphics[width=4.8cm]{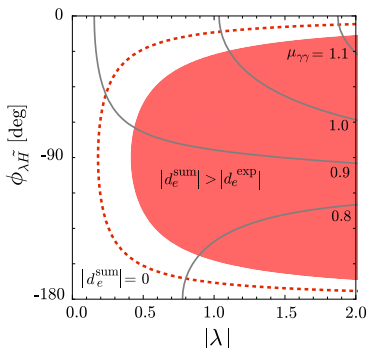} 
\caption{Impact of BAU-unrelated CP violation on $|d_e|$.
The region colored in red is excluded by the current experimental limit of the electron EDM, and the red dashed line corresponds to $|d^{\rm sum}_e|=|d^{WW}_e+d^{H\gamma}_e|=0$. 
The gray lines represent $\mu_{\gamma\gamma}=1.1,~1.0,~0.9$ and $0.8$ from top to bottom.
The input parameters are the same as in Fig.~\ref{fig:bau03_225}, but with
$m_{\tilde H}=300~{\rm GeV}$ and $m_{\tilde S}=277~{\rm GeV}$, which gives $Y_B/Y_B^{\rm obs}=1$.}
\label{fig:edm_sum}
\end{figure}

Next, we consider the case in which 
the BAU-unrelated CP-violating phase $\phi_{\lambda\tilde{H}}$
comes into play. 
Fig.~\ref{fig:edm_sum} shows the electron EDM in the $(|\lambda|, \phi_{\lambda\tilde{H}})$ plane. 
In this figure, we take the same input parameters as in Fig.~\ref{fig:bau03_225}, 
but with $m_{\tilde H}=300~{\rm GeV}$ and $m_{\tilde S}=277~{\rm GeV}$,
which yields $Y_B/Y_B^{\rm obs}=1$.
Here, we define $d_e^{\rm sum}=d_e^{WW}+d_e^{H\gamma}$.
Note that $d_e^{HZ}$ is accidentally suppressed with a factor 
of $(1/4-\sin^2\theta_W)\simeq0.02$ and thus numerically unimportant.
While the red region is excluded by the current limit of the electron EDM, 
the dashed line indicates the exquisite cancellation between $d_e^{WW}$ and $d_e^{H\gamma}$, resulting in $d_e^{\rm sum}=0$. 

In such a case, it is worth while to consider the other EDMs which might be complementary. The naive estimates show that $d_u\sim-1/3(m_u/m_e)d_e^{WW}$ and $d_d\sim2/3(m_d/m_e)d_e^{WW}$  under the condition of $d^{\rm sum}_e=0$, which lead to $d_n\sim d_p\sim {\cal O}(1)\times 10^{-28}~{e\cdot \rm cm}$. Although the current experimental bounds of $d_n$ and $d_p$ are not strong enough to probe this parameter region, the future experiments might be accessible \cite{npEDM}. Detailed analysis will be conducted in Ref. \cite{FHS}. 

Since $d_e^{H\gamma}$ is correlated with the signal strength of the Higgs decay to two gammas (denoted by $\mu_{\gamma\gamma}$, for the explicit formula, see, {\it e.g.,} Ref.~\cite{Chiang:2015fta}), we also examine it. $\mu_{\gamma\gamma}$ is represented by the gray lines:
$\mu_{\gamma\gamma}=1.1,~1.0,~0.9,$ and $0.8$ from top to bottom.
The whole region is still within the 2$\sigma$ region of the current LHC data,
$\mu_{\gamma\gamma} = 1.17\pm 0.27~(\text{ATLAS})$ and
$\mu_{\gamma\gamma} = 1.14^{+0.26}_{-0.23}~(\text{CMS})$.
We remark that the the sensitivity of $\mu_{\gamma\gamma}$ is expected to be 
improved up to $\mathcal{O}(5)\%$, 
and Higgs coupling to the gauge bosons ($\cos\gamma$ in the current setup) 
up to $\mathcal{O}(0.1)\%$ at future colliders
such as the high-luminosity LHC (HL-LHC)~\cite{HL-LHC},
International Linear Collider (ILC)~\cite{ILC} and TLEP~\cite{Gomez-Ceballos:2013zzn}. 
Therefore, the testability of EWBG in this scenario still persists.


\section{Conclusions}\label{sec:conclusion}
We have studied the relationship between the CP-violating source term for the BAU 
and the EDMs in the framework where the extra Higgs doublet and the singlet as well as
the new EW-interacting fermions ($\psi_{i,j}$) are introduced.
We scrutinized the ratio $\bar{S}_{\psi_i}$ (defined by Eq.~(\ref{Sbar})) 
as functions of the EW-interacting fermion masses. 
In the region where new fermions are degenerate, 
$\bar{S}_{\psi_i}$ is resonantly enhanced
due to the thermal effect appearing in the source term.
In the large mass limits of the fermions, on the other hand,
$\bar{S}_{\psi_i}$ gets milder or larger depending on the fermion species,
and the behaviors of which are mostly governed by the property of the loop function of the EDM rather than that of the CP-violating source term for the BAU.

As a concrete example, we considered the next-to-MSSM-like model 
and investigated the correlation between the BAU and the electron EDM for a typical parameter set.
It is found that 
as long as the BAU-related CP violation predominantly exists,
the current electron EDM places some constraints on 
the EWBG-favored region, and more importantly,  
it would probe the whole region if it is improved up to $1.0\times 10^{-29}~e\cdot{\rm cm}$.
However, once the BAU-unrelated CP violation comes into action, 
the strong connection between the BAU and electron EDM is not guaranteed any more,
which makes it challenging to probe the parameter space with the electron EDM only.
Nevertheless, even in such a case, the scenario could be probed with the aid of Higgs physics.

\begin{acknowledgments}
This work is supported by Grant-in-Aid for Scientific research from the Ministry of Education, Science, Sports, and Culture (MEXT), Japan, No. 23104011 (J.H.). The work of J.H. is also supported by World Premier International Research Center Initiative (WPI Initiative), MEXT, Japan.
The work of K.F. is supported by Research Fellowships of the Japan Society for the Promotion of Science for Young Scientists, No. 15J01079. E.S. is supported in part by the Ministry of Science and Technology, Taiwanunder Grant No. MOST 104-2811-M-008-011.
\end{acknowledgments}
%

\end{document}